\documentclass{kapedbk}
\usepackage{graphicx,amsmath,amsfonts,edbkps}
\normallatexbib

\newcommand{\tloc}{T_{\rm loc}}
\newcommand{\kf}{k_F}
\newcommand{\hc}{{\rm h.c.}}
\newcommand{\dback}{D_{\rm b}}
\newcommand{\dbacktilde}{{\tilde D}_{\rm b}}
\def \der#1{\frac{d#1}{dl}}
\newcommand{\xiloc}{\xi_{\rm loc}}
\newcommand{\fref}[1]{Fig.~\ref{#1}}

\newcommand{\normfig}{0.7\textwidth}
\newcommand{\largefig}{0.8\textwidth}

\begin{document}
\articletitle{Transport in Luttinger Liquids}

\chaptitlerunninghead{Transport in Luttinger Liquids}

\author{T. Giamarchi}
\affil{University of Geneva\\
1206 Geneva, Switzerland}
\email{Thierry.Giamarchi@physics.unige.ch}
\author{T. Nattermann}
\affil{Institut f\"ur Theoretische Physik, Universit\"at zu
K\"oln\\ Z\"ulpicher Str. 77 D-50937 K\"oln, Germany}
\email{natter@thp.uni-koeln.de}

\and 

\author{P. Le Doussal}
\affil{CNRS-Laboratoire de Physique Theorique de l'Ecole Normale
Superieure\\ 24 Rue Lhomond, Paris 75231 France.}
\email{Pierre.Ledoussal@physique.ens.fr}

\begin{abstract}
We compute the transport properties of one dimensional interacting
electrons, also known as a Luttinger liquid. We show that a
renormalization group study allows to obtain the temperature
dependence of the conductivity in an intermediate temperature
range. In this range the conductivity has a power-law like
dependence in temperature. At low temperatures, the motion proceed
by tunnelling between localized configurations. We compute this
tunnelling rate using a bosonization representation and an
instanton technique. We find a conductivity $\sigma(T) \propto
e^{-\beta^{1/2}}$, where $\beta$ is the temperature. We compare
this results with the standard variable range hopping (VRH)
formula.
\end{abstract}



\section{Introduction}

Since the discovery of Anderson localization
\cite{andersonlocalisation}, impurity effects in electronic
systems have always been a fascinating subject. Although our
understanding of the properties of noninteracting disordered
electronic systems is now rather complete
\cite{abrahamsloc,wegnerlocalisation,efetovlocalisation,efetovsupersymrevue},
the interacting case is still largely not understood. Indeed the
combined effects of disorder and interactions leads to a
reinforcement of both the disorder and interactions effects and
complicates greatly the physics of the problem
\cite{altshuleraronov,finkelsteinlocalizationinteractions,belitzlocalizationreview}.

One dimensional systems are an extreme realization of such a
situation. On one hand, even for noninteracting systems disorder
effects are extremely strong and all states are exponentially
localized \cite{berezinskiiconductivitylog,abrikosovrhyzkin}.
On the other hand for the pure system, interactions have an
extremely strong impact and lead to a non-fermi liquid state known
as a Luttinger liquid \cite{giamarchibook1d}. One can thus
expect a maximal interplay of disorder and interactions there.
However, in one dimension, good methods such as bosonization exist
to treat the interactions, so one can expect to have a more
complete solution even in presence of disorder.

We examine here the transport properties of such Luttinger
liquids.

\section{Model}

For simplicity we focuss here on spinless electrons. We consider
an interacting electronic system. Using the standard boson
representation \cite{giamarchibook1d} the Hamiltonian of such a
system is
\begin{equation} \label{eq:hambas}
 H = \frac1{2\pi} \int dx u K (\pi \Pi)^2 +
 \frac{u}{K}(\nabla\phi)^2 - \frac1{2\pi\alpha} \int dx \xi^*(x) e^{i 2\phi(x)}
\end{equation}
where $\xi(x)$ is a (complex) random potential representing the
backward (i.e. close to $2\kf$) scattering on the impurities.
$\xi(x)$ is taken to be gaussian and uncorrelated from site to
site
\begin{equation}
 \overline{\xi(x)\xi^*(x')} = \dback \delta(x-x')
\end{equation}
and all other averages are zero. The field $\phi$ is related to
the density of fermions by
\begin{equation} \label{eq:denslut}
 \rho(x) = -\frac1\pi\nabla\phi(x) + \frac1{2\pi\alpha} e^{i(2\kf x
 -2\phi)} + \hc
\end{equation}
and the current is simply $J = \partial_\tau\phi/\pi$.

In (\ref{eq:hambas}) the interaction effects among the electrons
are hidden in the two Luttinger parameters $u$ and $K$. $u$ is the
velocity of charge excitations. In the absence of interactions $u$
is the Fermi velocity $u = v_F$. $K$ is a dimensionless parameter,
controlling the decay of the various correlations. $K=1$ in the
absence of interactions and $K < 1$ for repulsive interactions.
$\alpha$ is a short distance cutoff of the order of the lattice
spacing.

An external electric field $E$ thus couples as
\begin{equation} \label{eq:elec}
 \int dx A J = -\frac1\pi\int dx E \phi(x)
\end{equation}
In the absence of disorder, the electric field makes the phase
$\phi$ grow with time. As can be seen from the Kubo formula, for
the pure system the conductivity is infinite. This corresponds
physically to the sliding of the electronic ``charge density
wave'' (\ref{eq:denslut}). Disorder pins the electronic density.
Such a pinning corresponds in the electronic language to the
Anderson localization of the electrons
\cite{giamarchicolumnarvariat,giamarchiquantumpinning}.

\section{Transport at intermediate temperatures}

The phase diagram of (\ref{eq:hambas}) has been extensively
studied and we refer the reader to
\cite{giamarchiloc,giamarchibook1d,glatzcdwfinitetemp} for
details. Renormalization group equations for the disorder $\dback$
can be written from the action
\begin{equation}\label{eq:incaction}
 S/\hbar = \int dx \;d\tau \left[ \frac1{2\pi K}\left[
 \frac1{u}(\partial_\tau \phi)^2 + u (\partial_x \phi)^2 \right] -
 \frac{\xi^*(x)}{2\pi\alpha\hbar}  e^{i 2 \phi(x)} + \hc \right]
\end{equation}
These equations are
\begin{equation}\label{rgk}
\begin{split}
 \der{K} &= - \frac{K^2}{2}\dbacktilde  \\
 \der{\dbacktilde}    &= (3-2 K)\dbacktilde \\
 \der{u} &=  - \frac{u K}{2} \dbacktilde
\end{split}
\end{equation}
where
\begin{equation} \label{eq:dbacktilde}
 \dbacktilde = \frac{2 \dback \alpha}{\pi u^2}
\end{equation}
For $K < 3/2$ the disorder is a relevant variable and leads to
localization. This of course includes the noninteracting point
$K=1$.

Using the RG, one can extract various physical quantities for the
disordered Luttinger liquid. For example, one can extract the
localization length. Let us renormalize up to a point where
$K^2\dbacktilde(l^*) \sim 1$. The true localization length of the
system is given by\index{localization length}
\begin{equation}
 \xiloc = e^{-l^*} \xiloc(l^*)
\end{equation}
but if $\dbacktilde(l^*) \sim 1$ the localization length of such a
problem is of the order of the lattice spacing. Thus,
\begin{equation}
 \xiloc \sim \alpha e^{-l^*}
\end{equation}
One can then integrate the flow to get $l^*$. The result depends
on the position in the phase diagram. When one is deep in the
localized phase (far from the transition) one can consider $K$ as
constant and thus
\begin{equation}
 \dbacktilde(l) = \dbacktilde(l=0) e^{(3-2K)l}
\end{equation}
Thus, \begin{equation} \label{eq:locdeep}
 \xiloc \sim \alpha \left(\frac{1}{K^2\dbacktilde}\right)^{\frac{1}{3-2K}}
\end{equation}

\begin{figure}[ht]
 \centerline{\includegraphics[width=\normfig]{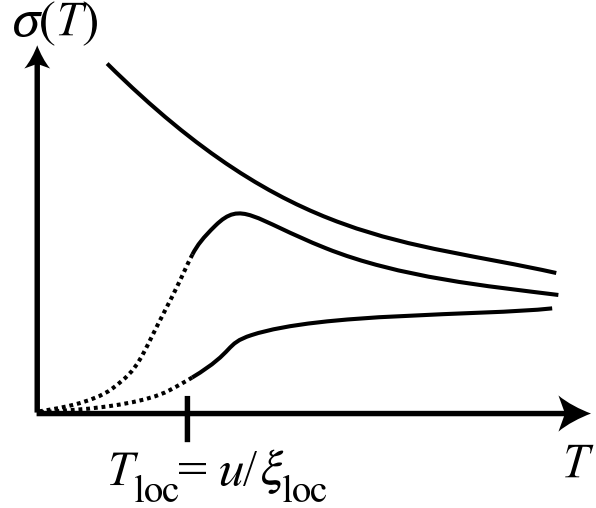}}
 \caption{Temperature dependence of the conductivity.
 For $K>3/2$ (top) the system is delocalized and the conductivity increases with
 decreasing temperature. For $1 < K < 3/2$ (middle) the system is localized but the
 conductivity starts increasing with decreasing temperature. The renormalization of
 $K$ due to disorder pushes the system to the localized side forcing the conductivity
 to decrease with decreasing temperature. For $K<1$ (bottom) the conductivity decreases with
 temperature even at high temperatures.  Below temperatures
 of the order of $u/\xiloc$, the system is strongly localized and the conductivity
 decreases exponentially (see text). This part (dashed line) cannot be extracted from the RG.}
 \label{fig:condtemp}
\end{figure}

One can also extract the frequency and temperature dependence of
the conductivity \cite{giamarchiloc}. Let us here look at the
temperature dependence by a very simple technique. The idea is
simply to renormalize until the cutoff is of the order of the
thermal length $l_T \sim u/T$ corresponding to $e^{l^*} \sim
l_T/\alpha$. At this lengthscale the disorder can be treated in
the Born approximation. As the conductivity is a physical quantity
it is not changed under renormalization and we
have\index{conductivity!disordered fermions}
\begin{eqnarray} \label{renodens}
 \sigma(n(0),\dbacktilde(0),0) = \sigma(n(l),D(l),l) = \sigma_0
 \frac{n(l)\dbacktilde(0)}{n(0)\dbacktilde(l)} = \sigma_0 \frac{e^l \dbacktilde(0)}{\dbacktilde(l)}
\end{eqnarray}
where $\sigma(n(l),\dbacktilde(l),l)=\sigma(l)$ and $n(l)$ are,
respectively, the conductivity and the electronic density at the
scale $l$. $\sigma_0=e^2 v_F^2/2\pi\hbar \dback$ is the
conductivity in the Born approximation, expressed with the initial
parameters. If one is deep in the localized phase, one can again
retain only the RG equation for the disorder and consider $K$ as
constant and one has
\begin{equation} \label{simpledis}
 \sigma(T) \sim \frac1{\dbacktilde} T^{2-2K}
\end{equation}
This result is schematized in \fref{fig:condtemp}.

\section{Creep}

Although one can use the RG to get the behavior of the
conductivity for $T > u/\tloc$, it cannot be used below this
energy scale since the flow goes to strong coupling. In order to
determine the transport properties at lower temperatures, we
compute the tunnelling rate between two static configurations of
the system. The details can be found in
\cite{nattermanntemperatureluttinger}, so we will recall here
only the main steps and results. We first use the RG equations to
reach a lengthscale at which the disorder $\dbacktilde$ becomes of
order one. This lengthscale corresponds to having a ``lattice
spacing'' that is now of the order of the localization length of
the system. Writing the disorder $\xi(x)$ as
\begin{equation}
 \xi(x) = |\xi(x)| e^{i2\zeta(x)}
\end{equation}
we now see that the disorder is minimized if on each ``site'' the
phase $\phi$ takes the value
\begin{equation} \label{eq:optimal}
 \phi(x) = \zeta(x) + \pi n_x
\end{equation}
where $n_x$ is an integer. The integer $n_x$ have to be chosen in
order to minimize the elastic term $(\nabla\phi(x))^2$ in
(\ref{eq:incaction}). Thus in the absence of the quantum term
$\Pi^2$ in the Hamiltonian the system is completely characterized
by the set of integer numbers $n_x$ \cite{glatzdiplom}. The
electric field (\ref{eq:elec}) wants to make the phase grow. Thus
in the presence of the quantum term in the action the phase will
tunnel between the optimal configurations described by
(\ref{eq:optimal}). This corresponds to an increase of $n_x$ by
one in some region of space. In order to compute the action
corresponding to such a tunnelling process one uses an instanton
technique as introduced in
\cite{colemaninstantonfield,makiinstantonperiodiclong} for
the pure system and \cite{nattermanntemperatureluttinger} in the
presence of disorder. The size $L_x$ in space and $L_\tau$ in time
of the instanton are determined by extremizing the action. One
finds
\begin{eqnarray} \label{eq:saddle}
 L^{\text{opt}}_x = \sqrt{\alpha/\epsilon} &\quad,\quad&
 L^{\text{opt}}_\tau = 1/(2\epsilon)
\end{eqnarray}
where
\begin{equation}
 \epsilon = \frac{2 K^* \alpha E}{\pi u^* \hbar} e^{2 l^*}
\end{equation}
and $^*$ denotes quantities as the scale $\xiloc$.

At zero temperature one can thus obtain the action of the
instanton as a function of the electric field. This leads to the
tunnelling rate
\begin{eqnarray} \label{eq:tunrate}
 P &\sim&  \text{exp}[- \frac1{\sqrt2}\left(\frac{\pi}{K^*}\right)^{3/2}\left(\frac{\Delta}{E \xiloc}\right)]
\end{eqnarray}
where we have introduced a characteristic energy scale $\Delta =
\hbar u^* /\xiloc$ associated with the localization length. Note
that $u^* /\xiloc$ is the pinning frequency
\cite{fukuyamapinning}. The expression (\ref{eq:tunrate}) leads
to a non-linear response. The linear conductivity is zero, and
such a process is the analogue of the creep for classical systems
\cite{blattervortexreview,nattermannvortexreview,giamarchivortexreview}.
In this case the system is able to overcome barriers by quantum
tunnelling, instead of thermal activation for the classical case.

For finite temperatures, the maximum size of the instanton in time
is $L_\tau < \beta \hbar u^* e^{-l^*}$. If the electric field
becomes too small the action associated with the tunnelling
process thus saturates. One thus recovers a linear response that
is given by
\begin{equation} \label{eq:vrhelas}
 \sigma(T) \propto e^{-\frac{S^*}{\hbar}} =
 \text{exp}[- \frac{\pi}{K^*}\sqrt{2\beta \Delta}]
\end{equation}

\section{Variable range hopping}

The expression (\ref{eq:vrhelas}) leads to the same temperature
dependence than the famous variable-range hopping law
\cite{mottmetalinsulator}. Let us briefly recall how the VRH law
is derived.

\begin{figure}[ht]
 \centerline{\includegraphics[width=\largefig]{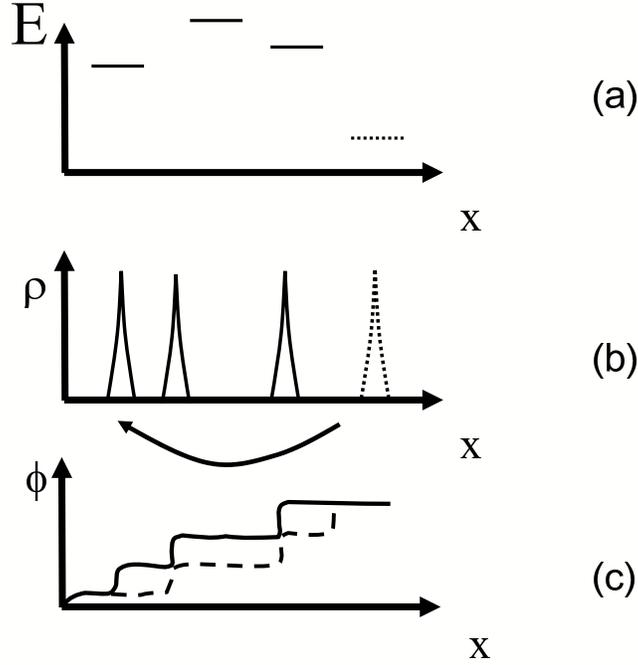}}
 \caption{(a) If the system is very localized, the eigenstates are localized in space $x$ over a distance of
 order $\xiloc$ and are spread in energy $E$. In order to transport current one should make a transition from an occupied state
 (dashed line) towards an empty state (full line). The difference in energy is provided by a coupling to a bath of phonons. This is the
 process at the root of the variable range hopping conductivity. (b) The density is a set of narrow peaks. Processes involved in the VRH thus
 transport charge from one localized state to another. (c) In the bosonized representation motion occur by shifting the phase by a multiple
 of $\pi$ in a finite region of space (whose size is determined by optimizing the action). Using the bosonization relations (see text) this
 corresponds precisely to the same transport of charge than in the VRH process.}
 \label{fig:vrh}
\end{figure}

The VRH law mostly applies to noninteracting electrons (or Fermi
liquids) in presence of phonons that can provide the inelastic
scatterings needed to make transitions between states of different
energies. One considers localized eigenstates at different
positions in space as indicated in \fref{fig:vrh}(a).
Thanks to phonons the system can make a transition from an
occupied state towards an unoccupied one. The probability of
making a transition involving a difference of energy $E$, between
two localized states at a distance $L$ is of order
\begin{equation}
 e^{-\beta E} e^{- L/\xiloc}
\end{equation}
Thus is it interesting to find transitions for states close to
each other. Of course such states are not in general close in
energy so there is a compromise between the energy and the
distance at which one can find states. If the density of states is
$N_0$, then the probability in a volume $L^d$ to find a state
within an interval of energy $E$ is
\begin{equation}
 N_0 L^d E
\end{equation}
thus a transition is possible when $N_0 L^d E \sim 1$. This leads
to a conductivity that is proportional to
\begin{equation}
 \sigma \propto e^{ - \frac{\beta}{N_0 L^d} - L/\xiloc}
\end{equation}
and optimizing with respect to $L$ one finds
\begin{equation}
 \sigma \propto e^{-(d+1)\left(\frac{\beta}{N_0 d^d \xiloc^d}\right)^{\frac{1}{d+1}}}
\end{equation}
In the presence of Coulomb interactions a similar formula
(Efros-Shklovskii) can be derived
\cite{efroscoulombgap,foglercoupledchainvrh} but with an
exponent $1/2$ instead of $1/(d+1)$. In one dimension both give
the same temperature dependence, but with different prefactors in
the exponential.

In fact the tunnelling process derived from the bosonized action
(\ref{eq:incaction}) is quite similar to the VRH process. Indeed
using the relation between the fermion density and the phase
(\ref{eq:denslut}) it is easy to see, as shown in
\fref{fig:vrh}(b) and (c) that a kink in $\phi$ corresponds to a
fermion at that position. The instanton corresponding to a shift
of $\phi$ over a finite region of space thus corresponds to moving
a particle from a localized state to another, as in the VRH
process.

The main difference is that in our approach the interactions are
totally treated. This is what gives the difference of prefactors
in the exponential. In our case the prefactors contains the
Luttinger liquid parameters (and thus the interactions). In the
case of VRH the prefactor simply depends on the density of states.
Our derivation provides an alternative derivation to VRH, directly
based on the bosonization representation and thus properly taking
into account the effects of interactions.

\section{Open issues}

Of course many open issues remain. In particular, both for the
intermediate temperature case and for the case of the creep, the
question of the precise role of dissipation arises. In the case of
the RG one assumes ,in order to obtain the conductivity, that the
temperature can be used as a cutoff. Doing so implicitly assumes
that the systems looses its coherence over a length $v/T$. This is
reasonable if the system is in contact with an external bath, but
is not the standard way to put the temperature in the Kubo
formula. Indeed normally the thermal bath is applied at time
$-\infty$ and then removed and all time evolution proceed simply
quantum mechanically. For the case of periodic potential the two
procedure are known to produce different results. If no phase
breaking processes are included the system being integrable has
too many conserved quantities and the conductivity remains
infinite at all finite temperatures
\cite{giamarchiumklapp1d,zotosconductivityintegrable,zotosconductivityheisenberg,roschconservation1d}.
This rather artificial result disappears in the more realistic
case where phase breaking processes are included
\cite{giamarchiumklapp1d,roschconservation1d}. For the
disordered case, in a similar way, the noninteracting system has a
conductivity that would remain zero at any finite temperature
since all states are exponentially localized, whereas the RG
result would gives a temperature dependence, as shown in
\fref{fig:condtemp}. In the creep regime similar questions arise.
In the VRH derivation the coupling to a phonon bath is explicitly
needed to provide the energy for the transition. In our analysis
such phonon bath is not needed, but on the other hand some
dissipation is required for the system to reach a steady state.
Whether the results we have found are valid as soon as an
infinitesimal dissipation is included, or whether
(\ref{eq:vrhelas}) would vanish with the dissipation in the limit
where the dissipation goes to zero is an important open question.

\begin{acknowledgments}
TG would like to thank B. L. Altshuler for interesting discussions
and the Swiss National Science foundation for support under MaNEP.
T. N. acknowledges financial support from the Volkswagen
foundation.
\end{acknowledgments}

\end{document}